\documentclass[%
 reprint,
 amsmath,amssymb,
 aps,
 prd,
floatfix,
showkeys
]{revtex4-1}

\usepackage[USenglish]{babel}
\usepackage{slashed}
\usepackage{graphicx}
\usepackage{dcolumn}
\usepackage{bm}
\usepackage{amssymb}
\usepackage{amsmath}
\usepackage{graphicx}
\usepackage{epsfig}
\usepackage{rotating}

\usepackage[utf8]{inputenc} 
\usepackage{amsmath} 
\usepackage{array}
\usepackage{float}
\usepackage{color}
\usepackage[usenames,dvipsnames]{xcolor}
\usepackage{epstopdf}
\usepackage{natbib}
\usepackage{graphicx}
\usepackage{lineno}
\usepackage{slashed}
\usepackage{color}
\usepackage{hyperref}
\usepackage{float}
\usepackage{mathtools,slashed}
\usepackage{multirow}

\newcommand{\be}{\begin{equation}}
\newcommand{\ee}{\end{equation}}
\newcommand{\bea}{\begin{eqnarray}}
\newcommand{\eea}{\end{eqnarray}}
\newcommand{\beas}{\begin{eqnarray*}}
\newcommand{\eeas}{\end{eqnarray*}}

\newcommand{\bse}{\begin{subequations}}
\newcommand{\ese}{\end{subequations}}

\begin{document}

\title{\bf Proton wave function in a water molecule: Breakdown of degeneration caused by interactions with the magnetic field of a Magnetic Resonance Imaging device}
\author{C. H.~Zepeda~Fern\'andez$^{1,2}$}
\email{Corresponding author, email:hzepeda@fcfm.buap.mx}
\author{J. L. Aguilar~Cuevas$^{2}$}
\author{E.~Moreno~Barbosa$^2$}

\address{
$^1$Cátedra CONACyT, 03940, CdMx Mexico\\
$^2$Facultad de Ciencias F\'isico Matem\'aticas, Benem\'erita Universidad Aut\'onoma de Puebla, Av. San Claudio y 18 Sur, Ciudad Universitaria 72570, Puebla, Mexico\\
}

\begin{abstract}
\begin{center}
\bf \large Abstract
\end{center}
The concept of a Magnetic Resonance Imaging (MRI) device is based on the emission of radio waves produced by the protons of the hydrogen atoms in water molecules when placed in a constant magnetic field after they interact with a pulsed radio frequency (RF) current. When the RF field is turned on, the protons are brought to a spin excited state. When the RF field is turned off, the MRI sensors are able to detect the energy released as the protons realign their spins with the magnetic field. In this work we provide a simple model to describe the basic physical mechanism responsible for the operation of MRI devices. We model the water molecule in terms of a central force problem, where the protons move around the (unstructured) doubly negatively charged oxygen atom. First, we employ an analytical treatment to obtain the system's wave function as well as its energy levels, which we show are degenerate. Next, the energy levels from the water molecule are studied in the presence of a uniform external magnetic field. As a result, they get shifted and the degeneration is lifted. We provide numerical results for a magnetic field strength commonly used in MRI devices. 
\end{abstract}


\maketitle

\section{Introduction}\label{1}

Magnetic Resonance Imaging (MRI) is a powerful technique to obtain three dimensional images of the structure and composition of human body parts~\cite{mr1,mr2,mr3}. The technique is not invasive, in the sense that it does not require the use of radiation. It is constantly evolving as the procedure to obtain the images improves~\cite{mri1,mri2,mri3,mri4}. An alternative technique to obtain three dimensional images is the Computed Tomography (CT). This technique is however invasive as it requires using x-rays~\cite{ct1,ct2,ct3} that can cause damage to body tissues~\cite{xr1,xr2,xr3}. In addition of not being invasive, MRI is preferred over CT because it can produce clearer images, particularly for the brain, the nerves, the muscles, the spinal cord, among other body parts.

The MRI main components are the magnet and the radio frequency coils (RFC)~\cite{mri}. MRI works by employing a strong magnetic field, usually in the range of 0.5-7 T, namely, six orders of magnitude larger than the magnetic field strength on the surface of the Earth. For this reason, patients must inform to MRI operators of any metallic implant they may have, as these can be affected by the strong magnetic field. A typical MRI procedure consists of laying down the patient within the device that contains the magnet. The magnetic field, interacts with the protons in the hydrogen atoms that make up the water molecules, abundantly present in the human body. The interaction is through the proton spin that can be oriented either parallel or anti parallel with respect to the magnetic field direction. A RF field produced by the RFC is used to reverse the orientation of the spin. When this RF is turned off, the  proton spin returns to their original orientation producing radio waves that are detected by the MRI device to produce the images~\cite{mri}. The human body is of course made out of different molecules and atoms and not only water. Moreover, the water molecule contains also protons and neutrons within the oxygen nucleus as well as electrons, all of which are subject to interactions with the magnetic field. It is therefore useful to describe why the most relevant interaction of the MRI magnetic field is the one with the protons in the hydrogen that makes up the water molecule. In this work we use a simple model to describe the basis of the interaction between the MRI magnetic field and the protons in the water molecule. The work is organized as follows: In Sec.~\ref{2} we first explain the reason to consider the spin of the protons in the water molecule to be the source of the main interaction for the MRI technique. In Sec.~\ref{3} we model the water molecule in terms of a central force problem and find the solution of the Schr\"odinger equation describing the protons wave function in the absence of a magnetic field. In Sec.~\ref{4} we introduce a constant magnetic field and find the breakdown of degeneration. In particular, we compute the shift of the energy levels with respect to the case in the absence of the magnetic field. We also compute the energy eigenvalues for typical field strengths employed in MIR devices. We finally discuss our results and conclude in Sec.~\ref{5}.

\section{Basic principles of the MRI technique}\label{2}

The human body is mainly composed of water. 80\% of human tissues consist of water molecules~\cite{water} which in turn are made of two hydrogen and one oxygen atoms. Let us first consider the protons in the hydrogen atoms. This is the principal assumption used for the physics in the MRI~\cite{PHMRI1,PHMRI2,PHMRI3}.  
Under normal circumstances, the orientation of the protons spins is arbitrary. When a human body is placed in a constant magnetic field, some proton spins in the water molecules become aligned whereas some others become anti-aligned with respect to the direction of the magnetic field. The latter correspond to excited energy states. If a RF is turned on, the effect is to reverse the spin alignment with respect to the magnetic field: parallel becomes anti-parallel and vice versa. When the RF is turned off and the proton spins return to the ground state, they release the excess energy in the form of electromagnetic radiation. 

Recall that the proton's Bohr magneton is given by
\begin{equation}
  \mu_{N_p}=\frac{e\hbar}{2m_p}=3.14\times10^{-8}{\rm eV/T},
  \end{equation}
and that the proton's Larmor frequency --a quantity which depends of the magnetic field strenght $B$-- is given by
\begin{equation}
\nu_p=\frac{2\mu_{p}B}{h},
\end{equation}
where $\mu_p=$2.79$\mu_{N_p}$ is the proton's magnetic moment, that depends on the sum of the magnetic moments of the quarks that make up the proton~\cite{mup}. The above means that, $\nu_p=42.595B$ MHz/T. Using a typical value for the field strength in a MRI device of $B=7$ T, one obtains that the protons emit radiation with a frequency $\nu_p=297.21$ MHz. In general, for typical values of the field strength used in these devices, this radiation is in the radio wave part of the spectrum.

The water molecule is also composed of electrons. Their corresponding Larmor frequency is given by $\nu_e=27.93 B$ GHz/T. Using $B=7$ T, one obtains that $\nu_e=195.53$~GHz. In general, for the typical field strengths used in MRI devices, $\nu_e$ is in the GHz range of the spectrum, namely in microwaves (MW). The other nucleons (including those from the oxygen atom) could also be subject to magnetic resonance~\cite{rf}. However, they are mostly spin-paired (their total spin is zero). This means in particular that they are not subject to a significant interaction with the magnetic field. Since the RFC is tuned to detect signals in the radio part of the spectrum, the MW radiation is signaled out. As a consequence, the magnetic field in MRI devices only detects transitions produced by the spin flip of protons in the water molecules.

After having spelled out these general remarks, we now proceed to study the proton wave function and energy levels when these are taken as being part of a water molecule. For simplicity we refer to these as the protons.

\section{Model for the water molecule}\label{3}

Finding analytical solutions of Schr\"odinger's equation for a system of many interacting particles (MIP) is nearly impossible. In most cases, this can be achieved only after resorting to some simplifying assumptions. The case of a water molecule is an example of a MIP system that has attracted considerable attention since long ago to find solutions using either suitable approximations~\cite{wm1,wm2,wm3,wm4} or numerical methods~\cite{wm5,wm6}.

As a first approximation, the water molecule is commonly described as having a di-polar structure, whereby the two protons of the hydrogen atoms represent the positive side, whereas the oxygen atom together with the two other electrons represent the negative side. Due to  electrostatic repulsion, these three components form a triangle~\cite{we1}. It is known that the distance $d_p$ between the protons of the two hydrogen atoms is, roughly speaking, constant, namely $d_p\sim$ 151.05~pm. On the same footing, the distance $d_o$ between the protons and the oxygen nucleus is about $d_o\sim$ 95.60~pm. One can then consider that the protons move as if they were a rigid body, with a fixed distance between them. The electrons of the two hydrogen atoms have a covalent bond with the oxygen atom. Therefore, the oxygen atom can be regarded as being a negative, doubly charged system that provides a central force for the motion of the two protons while its internal structure can be ignored, for the purpose of this study. The problem is then reduced to considering the motion of the two-proton system around a charged center with a charge  $Q=-2e$. 

\subsection{Hamiltonian }

The origin of the reference system is taken as the position of thee charge that produces the central force. As a first approximation the protons are considered as distinguishable. Since the distances between the protons and the protons and the oxygen nucleus are constant, the potential energy is also constant, yielding a system's Hamiltonian given by 
 \begin{eqnarray}
     \hat H_0 &=& \hat H_1 +\hat H_2\nonumber\\
     &=&\Big(\frac{\hat L_1^2}{2md_o^2}+\hat V_1(d_o,d_p)\Big)+\Big(\frac{\hat L_2^2}{2md_o^2}+\hat V_2(d_o,d_p)\Big),\nonumber\\
\end{eqnarray}
where 1 and 2 label the protons. The potential is given by
\begin{eqnarray}
    \hat V (d_o,d_p)&=&\hat V_1(d_o,d_p)+\hat V_2(d_o,d_p)\nonumber\\
    &=&\frac{1}{4\pi\epsilon_0}\Big(-\frac{2e^2}{d_o}+\frac{e^2}{2d_p}\Big)+\frac{1}{4\pi\epsilon_0}\Big(-\frac{2e^2}{d_o}+\frac{e^2}{2d_p}\Big)\nonumber\\
    &=&\frac{1}{4\pi\epsilon_0}\Big(-\frac{2e^2}{d_o}-\frac{2e^2}{d_o}+\frac{e^2}{d_p}\Big)\nonumber\\
    &=&\frac{1}{4\pi\epsilon_0}\Big(\frac{e^2}{d_p}-\frac{4e^2}{d_o}\Big).
\end{eqnarray}
Substituting the values for $d_p$ and $d_o$, we obtain $V=-23.4$ eV. 

\subsection{Protons wave function}

\begin{figure}[htbp]
\begin{center}
  \includegraphics[width=0.33\textwidth]{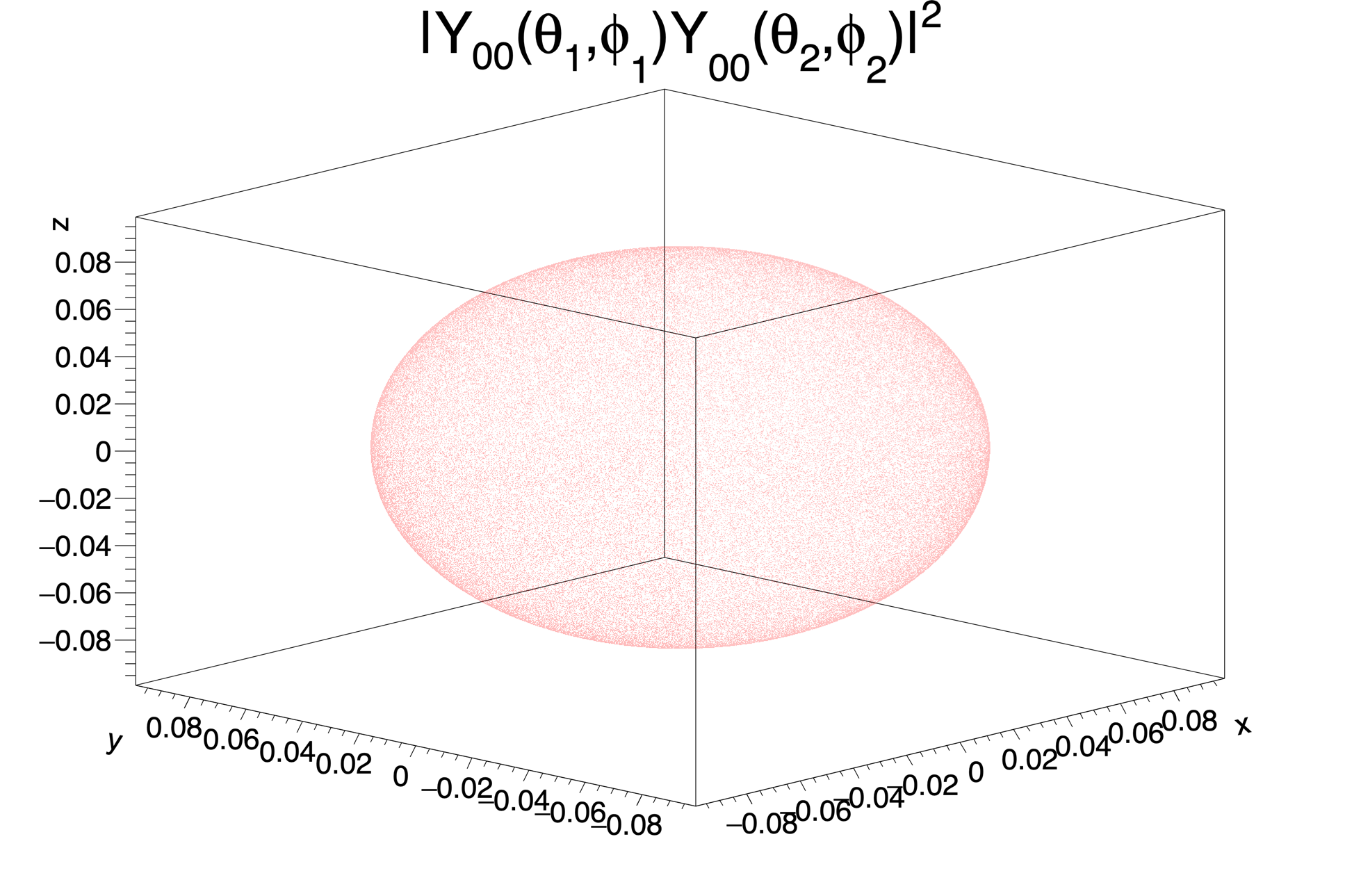}
  \includegraphics[width=0.33\textwidth]{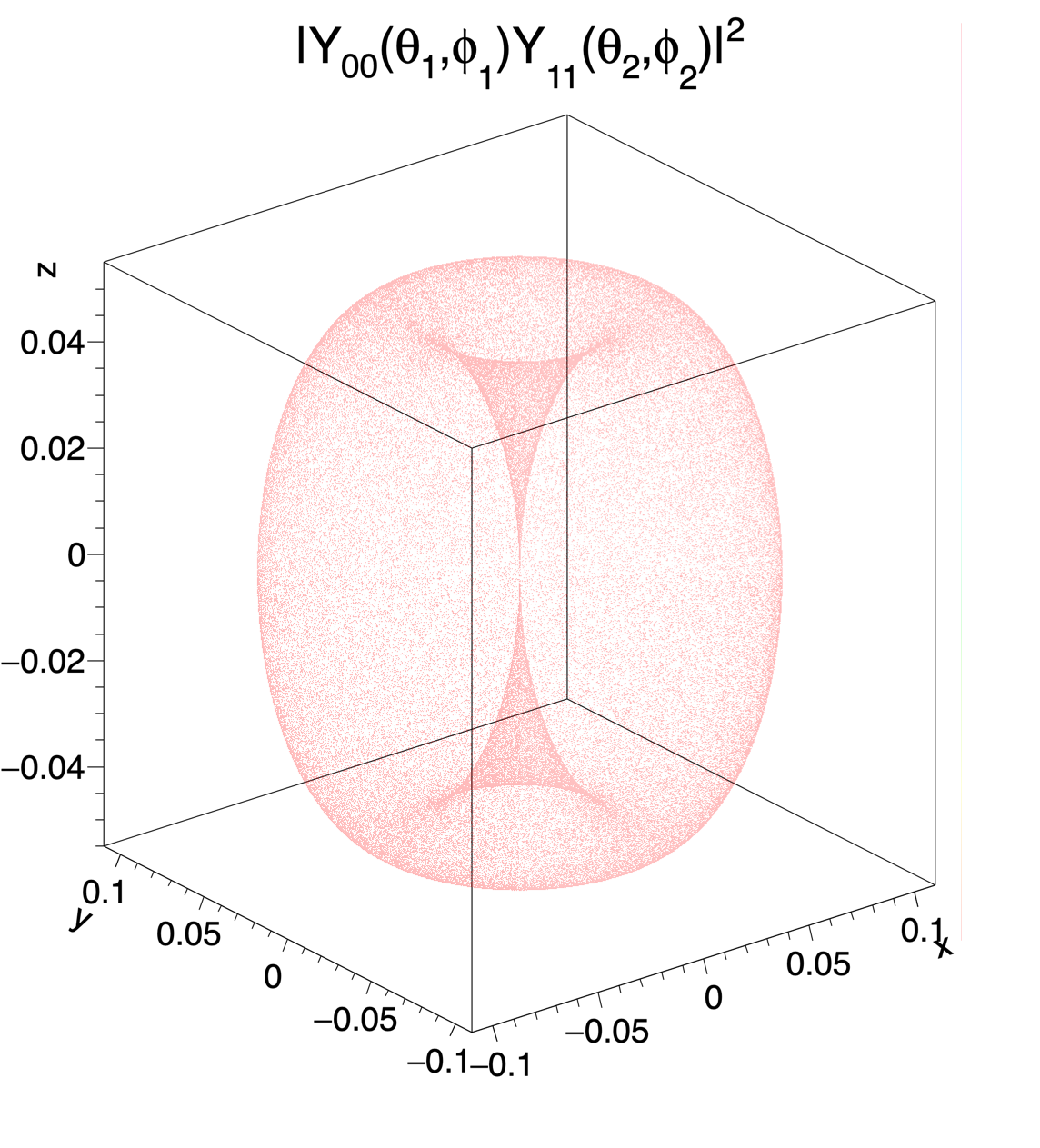}\\
  \includegraphics[width=0.33\textwidth]{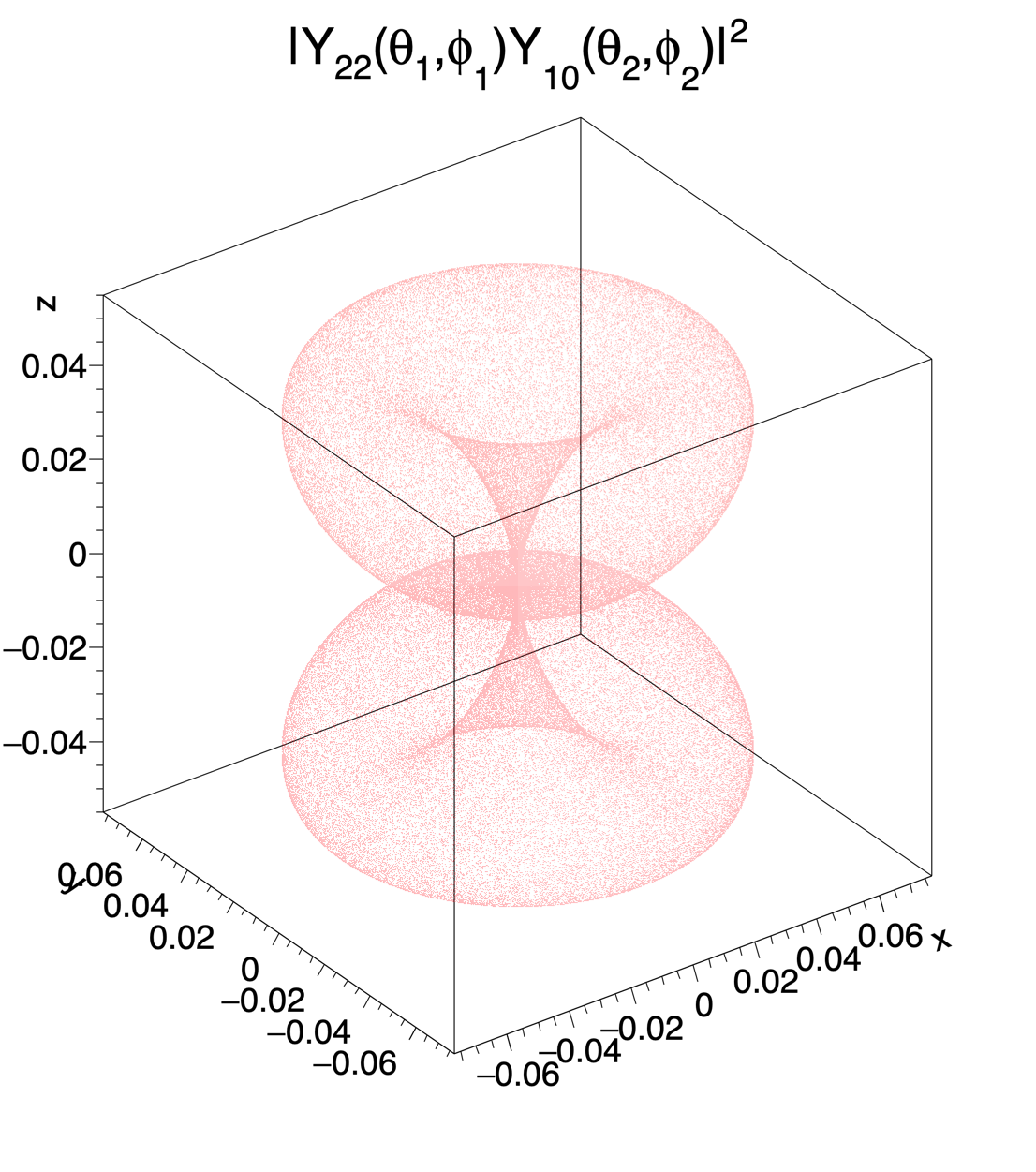}
  \includegraphics[width=0.33\textwidth]{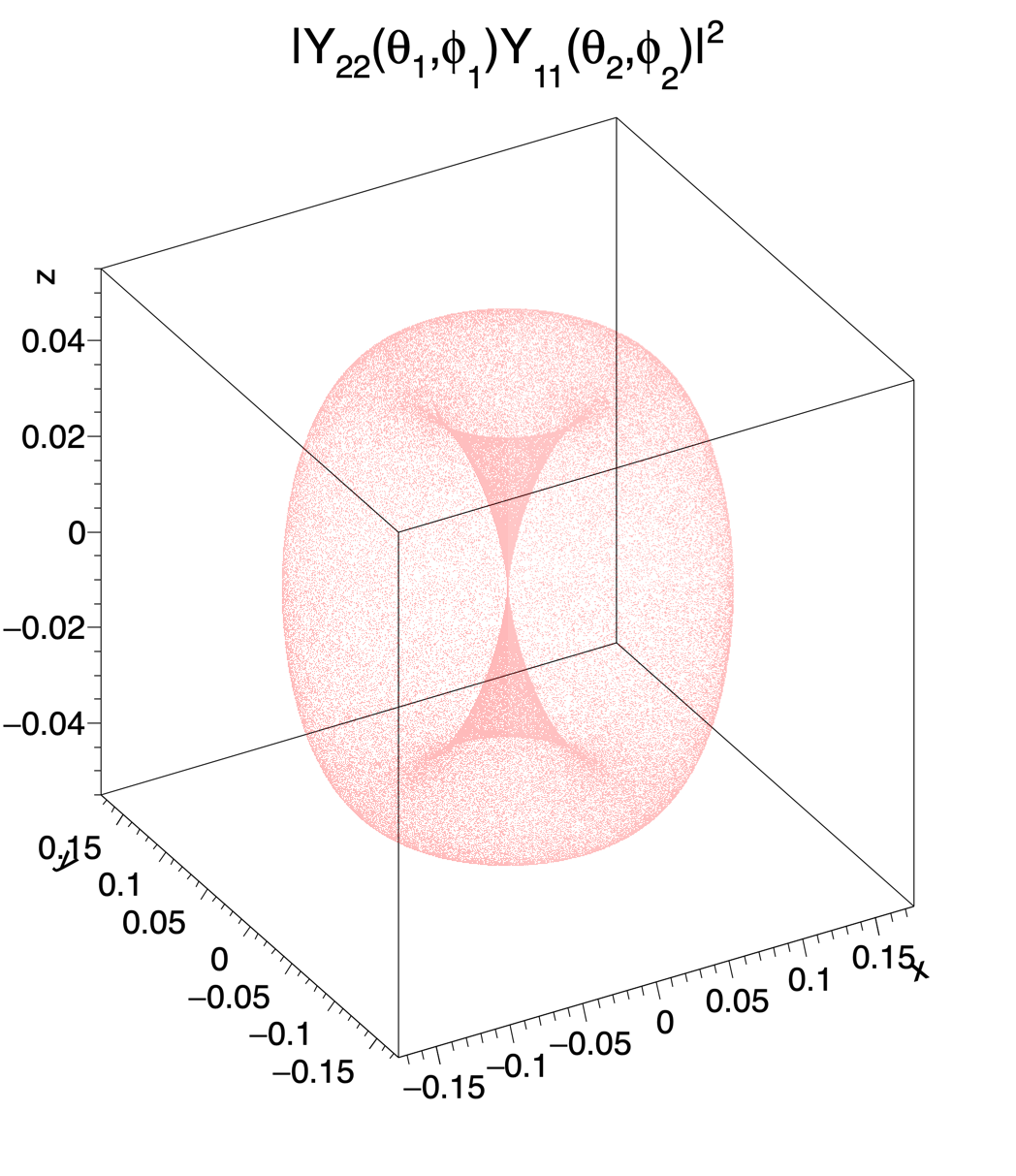}
\end{center}
\caption{Examples of probability distribution obtained from the corresponding wave functions of protons in the water molecule. The dimensions are arbitrary. }
\label{wave}
\end{figure}

Putting together all the above considerations, the Sch\"odinger's equation can be separated to be written as two equations, one for each proton, such that
\begin{equation} \label{p1}
  \Big[\frac{\hat L_1^2}{2md_o^2}+\hat V_1 \Big]\psi_1(\theta_1,\phi_1)=\epsilon_1\psi_1(\theta_1,\phi_1)
\end{equation}

  \begin{equation} \label{p2}
  \Big[\frac{\hat L_2^2}{2md_o^2}+\hat V_2 \Big]\psi_2(\theta_2,\phi_2)=\epsilon_2\psi_2(\theta_2,\phi_2),
\end{equation}
where $\hat L_i^2$ is the orbital angular momentum operator for the protons, $i=1, 2$. The total system's energy is given by $E=\epsilon_1+\epsilon_2$. Since $V<0$, then $\epsilon_i-V>0$ for both $i=1, 2$. As a  consequence, the solutions of Eqs.~(\ref{p1}) and~(\ref{p2}) represent the eigenfunctions of the orbital angular momentum, which are the Spherical harmonics, namely,
  \begin{equation} \label{sp1}
    \begin{split}
      \psi_1(\theta_1,\phi_1)=&Y_{l_1m_1}(\theta_1,\phi_1)\\
      =&(-1)^{m_1}\sqrt{\Big(\frac{2l_1+1}{4\pi}\Big)\frac{(l_1-m_1)!}{(l_1+m_1)!}}P_{l_1}^{m_1}(cos\theta_1)\\
      \psi_2(\theta_2,\phi_2)=&Y_{l_2m_2}(\theta_2,\phi_2)\\
      =&(-1)^{m_2}\sqrt{\Big(\frac{2l_2+1}{4\pi}\Big)\frac{(l_2-m_1)!}{(l_2+m_2)!}}P_{l_2}^{m_2}(cos\theta_2),
\end{split}
\end{equation}
where $P_{li}^{m_i}(\cos\theta_i)$ for $i=1,2$, are the associated Legendre polynomials. Therefore, the total system's eigenfunction is
\begin{equation} \label{sp}
\Psi_{l_1,m_1;l_2,m_2}(\theta_1,\phi_1;\theta_2\phi_2)=Y_{l_1m_1}(\theta_1,\phi_1)Y_{l_2m_2}(\theta_2,\phi_2),
\end{equation}
where $m_i=-l_i, -l_i+1,...,l_i-1,l_i$ for $i=1,2$. As an example, Fig.~\ref{wave} shows four probability distribution obtained from the corresponding proton wave functions in the water molecule.
The energy of the system is given by
\begin{equation}\label{E0}
E_{l_1,l_2}=\frac{\hbar^2}{2md_o^2}l_1(l_1+1)+\frac{\hbar^2}{2md_o^2}l_2(l_2+1)+V.
\end{equation}
We notice that for  given values of $l_1$ and $l_2$ the degeneracy is $(2l_1+1)\times(2l_2+1)$.

\subsection{Wave function for identical protons}

To finish the analysis of the wave function, as mentioned, the protons were treated as distinguishable. In order to formulate the model considering the protons as indistinguishable it is necessary to antisymmetrize the spatial part of the wave function, namely

\begin{equation}
  \begin{split}
  \Psi_{l_1,m_1;l_2,m_2}(\theta_1,\phi_1;\theta_2\phi_2)=\frac{1}{\sqrt 2}\Big[&Y_{l_1m_1}(\theta_1,\phi_1)Y_{l_2m_2}(\theta_2,\phi_2)\\
  \pm &Y_{l_2m_2}(\theta_1,\phi_1)Y_{l_1m_1}(\theta_2,\phi_2)\Big].
\end{split}
  \end{equation}

Since the proton sipin is $s=1/2$, they can be either in a singlet or a triplet state. Since the total wave function must be antisymmetric, when the spatial wave function is symmetric, the spin part must be antisymmetric (singlet) whereas if the spatial wave function is antisymmetric, the spin part is symmetric (triplet).

\section{Water molecule in a magnetic field}\label{4}

To introduce the effects of a magnetic field ${\bf B}$, we make the minimal substitution ${\bf p} \rightarrow {\bf p}-\frac{q}{c}{\bf A}$, where ${\bf A}$ is the magnetic potential and ${\bf B}=\nabla \times {\bf A}$. Therefore, the Hamiltonian for each proton becomes
\begin{equation}
  {\bf \hat H}_i=\frac{1}{2m}\big({\bf \hat p_i}-\frac{q}{c}{\bf \hat A}\big)^2; \hspace{1cm} i=1,2,
\end{equation}
resulting in the total system's Hamiltonian 
\begin{eqnarray}
  {\bf \hat H}&=&\frac{1}{2m}\big({\bf \hat p}_1-\frac{q}{c}{\bf \hat A}\big)^2+\frac{1}{2m}\big({\bf \hat p}_2-\frac{q}{c}{\bf \hat A}\big)^2 + V\nonumber\\
  &= & {\bf \hat H_0} -\frac{q}{2mc}( {\bf \hat A}\cdot {\bf \hat p_1}+ {\bf \hat p_1}\cdot {\bf \hat A})\nonumber\\
  &-&\frac{q}{2mc}( {\bf \hat A}\cdot {\bf \hat p_2}+ {\bf \hat p_2}\cdot {\bf \hat A}) +\frac{q^2}{mc^2}{\bf \hat A}^2,
\end{eqnarray}
where  ${\bf \hat H_0}$ is the Hamiltonian with $B=0$. Working in the Coulomb gauge $\nabla \cdot {\bf \hat A} =0$, one obtains that $ {\bf \vec p_i} \cdot {\bf \hat A} -{\bf \hat A}\cdot {\bf \hat p_i}=0$. Finally, using that ${\bf \hat A}=\frac{1}{2}{\bf B}\times {\bf \hat r}$ and ${\bf \hat L} ={\bf \hat r}\times {\bf \hat p}$, the Hamiltonian is
given by
\begin{equation}\label{HB1}
  \hat H=\hat H_0-\frac{q}{2mc}({\bf B} \cdot {\bf \hat L_1})-\frac{q}{2mc}({\bf B} \cdot {\bf \hat L_2})+\frac{q^2}{mc^2}{\bf A}^2.
\end{equation}
Since the magnetic field is directed along the $\hat{z}$-axis Eq.~(\ref{HB1}) reduces to
\begin{equation}\label{HB}
  \hat H=\hat H_0-\frac{q}{2mc}(B \cdot \hat L_{z1})-\frac{q}{2mc}(B \cdot \hat L_{z2})+\frac{q^2}{4mc^2} (x^2+y^2)
\end{equation}
The last term can be neglected as it is inversely proportional to $c^2$, even when $B$ is strong. Therefore the Hamiltonian is given by
\begin{equation}\label{HB2}
  \hat H=\hat H_0-\frac{qB}{2mc}( \hat L_{z1}-\hat L_{z2})
\end{equation}

\begin{figure}[htbp]
\begin{center}
\includegraphics[width=0.5\textwidth]{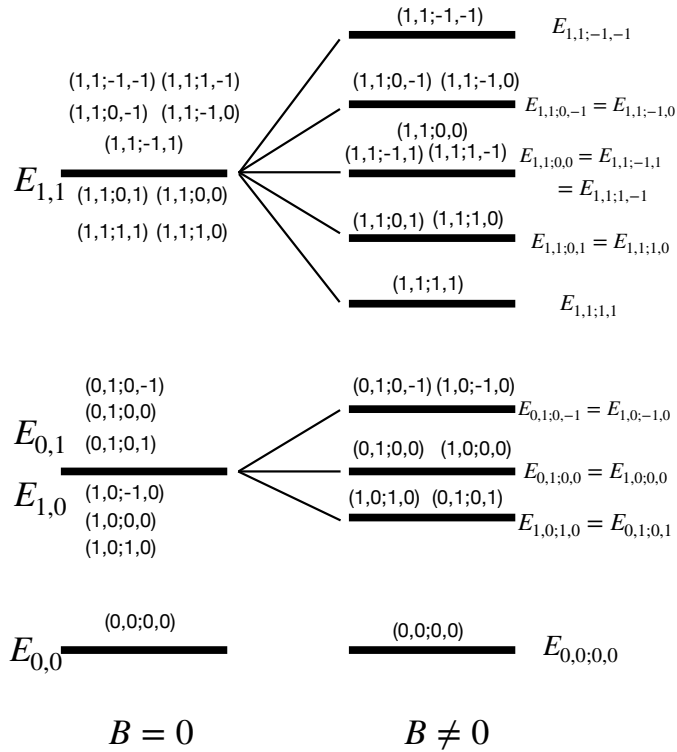}
\end{center}
\caption{Left: For the $B=0$ case, the energy levels are degenearted in the quantum numbers $l_1,l_2,m_1,m_2$. Right: For the $B\neq0$ case, the degeneracy with respect to the quantum numbers $m_1, m_2$ is lifted. The system is however still degenerated with respect to $l_1, l_2$. The notation $(l_1,l_2;m_1,m_2)$ represents the quantum numbers associated to the wave function $\Psi_{l_1,m_1;l_2,m_2}$.}
\label{break}
\end{figure}

It is easy to note that $\hat H$, $\hat H_0$, $\hat L_{z1}$ and $\hat L_{z2}$ commute with each other. Therefore, the common set of eigenfunctions is given by $|l_1m_1l_2m_2>=\Psi_{l_1,m_1;l_2,m_2}$. Using this basis, one can compute the energy eigenvalues
\begin{equation}\label{EB}
  \begin{split}
  E_{l_1,m_1;l_2,m_2}=&<l_1m_1l_2m_2|\hat H|l_1m_1l_2m_2>\\
  =&E_{l_1,l_2}-m_1\hbar B\mu_p-m_2\hbar B\mu_p\\
  =&\frac{\hbar^2}{2md^2}[l_1(l_1+1)+l_2(l_2+1)]+V\\
  &-m_1 \hbar B\mu_p-m_2\hbar B\mu_p,
  \end{split}
  \end{equation}
where, $\mu_p=\frac{e}{2mc}$. Defining $\omega_p=\mu_pB$ the energy can be written as
\begin{equation}\label{ew}
  \begin{split}
  E_{l_1,m_1;l_2,m_2}=&\frac{\hbar^2}{2md^2}[l_1(l_1+1)+l_2(l_2+1)]+V\\
  &-m_1 \hbar\omega_p-m_2\hbar\omega_p\\
  =&E_{l_1,l_2}-m_1 \hbar\omega_p-m_2\hbar\omega_p
    \end{split}
  \end{equation}
Notice that the degeneracy in the quantum numbers $m_1$ and $m_2$, obtained for the $B\neq0$, is now lifted. When $B=0$, $\omega_p=0$, therefore the energy eignevalues for the vanishing magnetic field case, $E_{l_1,l_2}$, are recovered. As an example, Fig.~\ref{break} shows the first two energy levels comparing the cases when $B=0$ and $B\neq0$.

In a MRI study, the magnetic field strength can be anywhere in the range of 0.5~T and 7~T. Using Eq.~(\ref{ew}), it is simple to compute the proton energy levels for any magnetic field strength. As an example, we take $B=7$~T. Table~\ref{table} shows the values of the energy levels of the protons in the water molecule. Shown are the values for the ground state and the first and second excited states. We notice that the degeneracy in the quantum numbers $m_1$ and $m_2$ is removed,  as illustrated also in Fig.~\ref{break}.

\begin{table}[htbp]
\caption{Energy levels for the ground, first and second state of the protons in the water molecule.}
\label{table}
\centering
\smallskip
\begin{tabular}{| c | c | c |}
\hline
State & ($l_1,l_2;m_1,m_2$) & Energy (eV))   \\
\hline
\hline
Ground & (0,0;0,0) & -23.4  \\
\hline
\hline
\multirow{3}{1cm}{First} & (0,1;0,-1)=(1,0;-1,0) & -23.39545379 \\ \cline{2-3}
 & (0,1;0,0)=(0,1;0,0)  &  -23.397727 \\ \cline{2-3}
& (1,0:1,0)=(0,1;0,1) & -23.39545421\\ 
\hline
\hline
\multirow{3}{1cm}{Second} & (1,1;-1,-1) & -23.39090758\\ \cline{2-3}
 & (1,1;0,-1)=(1,1;-1,0) & -23.39090779\\ \cline{2-3}
 & (1,1;0,0)=(1,1;-1,1)=(1,1;1,-1) & -23.390908\\ \cline{2-3}
 & (1,1;0,1)=(1,1;1,0) & -23.39090821\\ \cline{2-3}
 & (1,1;1,1) & -23.39090842\\ \cline{1-3}

\end{tabular}
\end{table}

\section{Discussion and conclusions}\label{5}
The working principle behind MRI devices is the interaction of the constant magnetic field with the spin of the protons (hydrogen nuclei) in the water molecule, we worked on this idea to know and understand their energy levels. By using a simple model to describe the water molecule, we formulated and analytically solved the Sch\"odinger's equation. In this model, since the protons are described as having a constant separation between them and with respect to the oxygen nucleus, they move as a rigid body around the oxygen atom, which is regarded as a double negatively charged source of a central force. Notice that the result of this simple description of the interaction between the magnetic field and the protons in the water molecule is similar to the Zeeman effect, where a lifting of the degeneration is also induced by the magnetic field. The similarity comes from the model of the water molecule that we used, which is the analog to the hydrogen atom system with two particles interacting in a central potential. 
The wave function for the case of $B=0$ is a product of the spherical harmonics, and thus it is described by the four quantum numbers $l_1,~l_2,~m_1$ and $m_2$. We have obtained the explicit energy degeneracy for all quantum numbers. This degeneracy is lifted when the water molecule is immersed in a constant magnetic field. However, the degeneracy persists with respect to the quantum numbers $l_1$ and $l_2$. Comparing with the case of the anomalous Zeeman effect, it is expected that this last degeneracy is lifted when the spin of both protons is also accounted for.\\
With this analysis, we show the energy levels for the protons interacting with a magnetic field, hoping, these results can be used to improve the techniques to obtain the image from a MRI, having more finesse.

\section*{Acknowledgments}
The authors thank A. Ayala for a thorough reading and suggestions to improve this work. 

\begin{minipage}{.5\textwidth}
\centering
\bf \large References
\end{minipage}


\begin{thebibliography}{13}
\bibitem{mr1}
 Care at  Mayo Clinic. \url{https://www.mayoclinic.org/tests-procedures/mri/about/pac-20384768}.
\bibitem{mr2}
  Web MD. \url{https://www.webmd.com/a-to-z-guides/what-is-an-mri}.
\bibitem{mr3}
  Medical News Today \url{https://www.medicalnewstoday.com/articles/146309}.
\bibitem{mri1}
  J. A. Fessler. {\textit Optimization Methods for Magnetic Resonance Image Reconstruction: Key Models and Optimization Algorithms}, IEEE Signal Processing Magazine, {\bf 37} 1, pp. 33-40 (2020). \url{doi: 10.1109/MSP.2019.2943645}.
\bibitem{mri2}
  F. Knoll et. al. {\textit Deep Learning Methods for Parallel Magnetic Resonance Image Reconstruction}. 
  arXiv:1904.01112v1 [eess.SP].
\bibitem{mri3}
  Xiaobo Qu, et al. {\textit Magnetic resonance image reconstruction from undersampled measurements using a patch-based nonlocal operator}. Medical Image Analysis {\bf 18} 6 pp 843-856. \url{https://doi.org/10.1016/j.media.2013.09.007}.
\bibitem{mri4}
  B. Wen, et al. {\textit Transform Learning for Magnetic Resonance Image Reconstruction: From Model-Based Learning to Building Neural Networks}, IEEE Signal Processing Magazine {\bf 37} 1 pp. 41-53 (2020)  \url{doi:10.1109/MSP.2019.2951469}.
\bibitem{ct1}
  Mayo Clinic. \url{https://www.mayoclinic.org/tests-procedures/ct-scan/about/pac-20393675}
\bibitem{ct2}
  National Institute of Biomedical Imaging and Bioengineering. \url{https://www.nibib.nih.gov/science-education/science-topics/computed-tomography-ct}.
\bibitem{ct3}
  RaiologyInfo.org For patients. \url{https://www.radiologyinfo.org/en/ctscan}.
\bibitem{xr1}
  Medical News Today. \url{https://www.medicalnewstoday.com/articles/219970}.
\bibitem{xr2}
  Harvard Health Publishing Harvard Medical School. \url{https://www.health.harvard.edu/cancer/radiation-risk-from-medical-imaging}.
\bibitem{xr3}
  U.S. Food \& drugs Administration. \url{https://www.fda.gov/radiation-emitting-products/medical-imaging/medical-x-ray-imaging}.
\bibitem{mri}
  S.D. Serai, et al. Components of a magnetic resonance imaging system and their relationship to safety and image quality. Pediatr. Radiol. {\bf 51} 5. 716-723 (2021). \url{doi: 10.1007/s00247-020-04894-9. Epub 2021 Apr 19. PMID: 33871725}.
\bibitem{water}
Zhang J.et al. The Translationally Controlled Tumor Protein and the Cellular Response to Ionizing Radiation-Induced DNA Damage: 12.2.1 Direct and Indirect Effects of Ionizing Radiation (2017).
\bibitem{PHMRI1}
R. A. Pooley. Fundamental Physics of MR Imaging. The AAPM/RSNA physics tutorial for residents {\bf 25} 4 (2005) 1087-1099. \url{https://doi.org/10.1148/rg.254055027}. 
\bibitem{PHMRI2}
A. Pai , R. Shetty , Y.S. Chowdhury. Magnetic Resonance Imaging Physics. StatPearls (2021)  \url{https://www.ncbi.nlm.nih.gov/books/NBK564320/}
\bibitem{PHMRI3}
Vassilios S Vassiliou et. al. Magnetic resonance imaging: Physics basics for the cardiologist. JRSM Cardiovascular Disease. {\bf 7} I-9 (2018). \url{DOI: 10.1177/2048004018772237}.
\bibitem{mup}
J.H. Sanders and K. C. Turberfield. The Magnetic Moment of the Proton. I. The Value in Nuclear Magnetons. Proc. R. Soc. Lond. {\bf 272} A 79-102 (1963). \url{doi: 10.1098/rspa.1963.0043}.
\bibitem{rf}
  Britannica, The Editors of Encyclopaedia. Magnetic resonance. Encyclopedia Britannica, 25 Jan. 2011, \url{https://www.britannica.com/science/magnetic-resonance}. Accessed 14 September 2021.
\bibitem{wm1}
  R. McWeeny and K. A. Ohno {\textit A quantum-mechanical study of eater molecule}. Proc. R. Soc. Lond. {\bf 255} A. pp. 367-381 (1960). \url{doi: 10.1098/rspa.1960.0072}.
\bibitem{wm2}
  J. Chem. {\textit Multiconfiguration Wavefunctions for the Water Molecule}. J. Chem. Phys. {\bf 55} 1720 (1971). \url{https://doi.org/10.1063/1.1676302}.
\bibitem{wm3}
  A. Sprague Coolidge. {\textit A Quantum Mechanics Treatment of the Water Molecule}. Phys. Rev. {\bf 42} 189 (1932).
\bibitem{wm4}
  Soe Aung. Thesis of Doctor of Philosophy. {\textit Part I Approximate Hartree-Fock Wavefunctions One-electron Properties and Electronic Structure of the W ater Molecule. Part II Perturbation-Variational Calculation of the Nuclear Spin-Spin Isotropic Coupling Constant in HD} California Institute of Technology Pasadena California (1969).
\bibitem{wm5}
B. Clark, et. al. {\textit Computing the energy of a water molecule using MultiDeterminants: A simple, efficient algorithm}. J. Chem. Phys. {\bf 135} 244105 (2011). \url{doi:10.1063/1.3665391}
\bibitem{wm6}
  M. Johansson, {\textit Introduction to Electronic Structure Theory}. CSC/PRACE Spring School in Coputational Chemistry 2018. \url{http://iki.fi/mpjohans}.
\bibitem{we1}
  S. S. Zumdahl. {\textit water} Encyclopedia Britannica. (2021). \url{https://www.britannica.com/science/water}.
\end{thebibliography}
\end{document}